\documentstyle[twoside,fleqn,espcrc2,epsfig]{article}

\newenvironment{mydescription}[1]%
 {\begin{list}{}{%
  \setlength{\leftmargin}{10pt}%
  \setlength{\itemindent}{10pt}%
  \setlength{\rightmargin}{0pt}}}
  {\end{list}}

\title{Simulating 4D Simplicial Gravity including Degenerate 
       Triangulations}
\author{
  {\bf S.~Bilke}\address{Inst.~Theor.~Fysica, Univ.~Amsterdam, 
                   1018 XE Amsterdam, The Netherlands \\
		   $^{\rm b}$Fakult\"{a}t f\"{u}r Physik, Universit\"{a}t 
                   Bielefeld, 33501 Bielefeld, Germany }
  and G.~Thorleifsson$^{\rm b}$
}		             

\begin{document}

\begin{abstract}
We extend simulations of simplicial gravity in four dimensions
to include {\it degenerate} triangulations and 
demonstrate that using this ensemble the
geometric finite-size effects are much reduced.
We provide strong numerical evidence for the existence 
of an exponential bound on the entropy of the model
and establish that the phase structure
is identical to that of a corresponding model restricted
to an ensemble of combinatorial triangulations. 
\end{abstract}

\maketitle

\section{INTRODUCTION}

In discretized models of four-dimensional Euclidean quantum gravity, 
known as simplicial gravity, the integration over metrics is
replaced by summations over an ensemble of triangulations constructed by all
possible gluings of equilateral 4-simplexes into closed (piece-wise linear)
simplicial manifolds (see e.g.\ Ref.~\cite{janbook}). The regularized
Euclidean Einstein-Hilbert action is particularly simple in this approach;
it can be taken to depend on only two coupling constants, $\kappa $ and $\mu 
$, related to the inverse Newton's and cosmological constants. The
regularized grand-canonical partition function thus becomes: 
\begin{equation}
 Z(\mu ,\kappa )\;=\;\sum_{T\in \mathcal{T}}\;\frac{1}{C_{T}}\;
 {\rm e}^{ \; \textstyle -\mu N_{4}+\kappa N_{2}}.  
 \label{model}
\end{equation}
The sum is over all distinct triangulations $T\in \mathcal{T}$, $N_{i}$ is
the number of $i$-simplexes in the triangulation $T$ and $C_{T}$ denotes its
symmetry factor --- the number of equivalent labelings of the vertexes.

Extensive numerical simulations have established that the model 
Eq.~(\ref{model}) has a strong-coupling (small $\kappa $) 
crumpled phase and a
weak-coupling (large $\kappa$) elongated phase, 
separated by a discontinuous phase
transition. In the crumpled phase the geometry is dominated by a singular
structure; two singular vertexes connected to an extensive fraction of the
total volume, joined by a sub-singular edge.  The
elongated phase, on the other hand, is dominated by essentially
one-dimensional (tree-like) triangulations  --- branched polymers. 

\section{DEGENERATE TRIANGULATIONS}
In Eq.~(\ref{model}), $\mathcal{T}$ denotes a suitable ensemble of
triangulations included in the partition function. Different ensembles are
defined by imposing various restriction on how the simplexes are glued
together. Provided this leads to a well-defined partition function, and as
long as this difference is only at the level of discretization, one expects
different choices of $\mathcal{T}$ to lead to the same continuum theory in
the thermodynamic limit. This is known to be true in two dimensions where
models of simplicial gravity corresponding to different choices of $\mathcal{%
T}$ are soluble as matrix models \cite{mat2d}. 

All simulations of four-dimensional simplicial gravity have, as of today, used
an ensemble of \textit{combinatorial} triangulations $\mathcal{T}_C$. In a
combinatorial triangulation every 4-simplex is uniquely defined
by a set of five {\it distinct} vertexes --- it is said to be
combinatorially unique. Here we report on simulations of
the model Eq.~(\ref{model}) defined with a larger
ensemble $\mathcal{T}_D$ including \textit{degenerate} triangulations.
We relax the above constraint and allow distinct simplexes to be defined 
by the same set of vertexes. We do, however, retain the restriction that every 
4-simplex is defined by a set of distinct vertexes, i.e.\ we exclude
degenerate simplexes. Clearly $\mathcal{T}_C \subset \mathcal{T}_D$. 

The benefit of using a larger ensemble of triangulations is well known
from simulations of simplicial gravity in two \cite{fss2d} and 
three \cite{deg3d} dimensions.  It is established 
that less constrained the triangulations are the smaller
the geometric finite-size effects are.
Simulations of four-dimensional simplicial gravity are 
notoriously time-consuming, 
primarily due to the large volumes needed to observe any ''true`` infinite 
volume behavior, hence any reduction in the finite-size effects is 
of great practical importance.

\section{RESULTS}
We have simulated the model Eq.~(\ref{model}) using degenerate
triangulations on volumes up to 25.600 4-simplexes using Monte Carlo
methods. As customary we work in a quasi-canonical ensemble of spherical
manifolds with almost fixed $N_{4}$: 
\begin{equation}
Z(\mu ,\kappa ;\bar{N}_{4})\;=\;\sum_{N_{4}}{\rm e}^{-\mu N_{4}-\delta
(N_{4}-\bar{N}_{4})^{2}}\;\Omega _{N_{4}}(\kappa ),  \label{pseudo}
\end{equation}
\vspace{-3pt}
where $\Omega _{N_{4}}(\kappa )=\sum_{T\in {\cal T}(N_{4})}\exp (\kappa
N_{2})$ is the canonical partition function. As there do not exist ergodic
volume conserving geometric moves, the canonical ensemble cannot be
simulated directly, we must allow the volume to fluctuate. The quadratic
potential term added to the action ensures, for an appropriate choice of $%
\delta $, that these fluctuations are small.

In the simulations the triangulation-space is explored using a set of local
geometric changes, the ($p,q$)--moves \cite{moves}. For
combinatorial triangulations the ($p,q$)--moves are known to be ergodic for 
$\mathrm{D} \leq 4$. To demonstrate that the same holds true for
degenerate triangulations we observe that every set of combinatorially
equivalent simplexes, or sub-simplexes, can be made distinct by a
finite-sequence of the ($p,q$)--moves. Thus every degenerate triangulations
can be reached from a combinatorial one. In addition, the local nature of the 
($p,q$)--moves
prohibits the creation of pseudo-manifolds in the simulations, i.e.\
triangulations containing vertexes with a neighborhood not homeomorphic to
the 4-sphere.

\begin{figure}[t]
\centerline{\epsfig{file=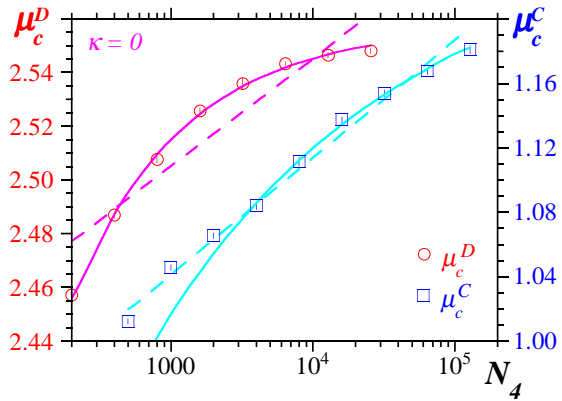,width=2.8in}}
\caption{The pseudo-critical cosmological constant, $\protect\mu_c^D(N_4)$,
 together with fits to power-law convergence (solid) 
 and to logarithmic divergence (dashed).  Included are the corresponding
 values, $\protect\mu_c^C(N_4)$, for combinatorial triangulations.}
\label{fig1}
\end{figure}
\begin{table}[tbp]
\vspace{-8pt}
\caption{Parameters in fits of $\protect\mu_c^D(N_4)$ 
 to Eqs.~(\ref{exp}) and (\ref{super}), including volumes
 $N_4 = 400$ to 25.600.}
\begin{center}
\begin{tabular}{lllc} \hline
& $\bar{\mu}$ & $\gamma$ & $\chi^2$/(d.o.f). \\ \hline
Eq.~(\ref{exp}) & 2.556(3) & 0.55(5) & 3.8 \\ 
Eq.~(\ref{super}) & 2.385(4) & & 117\\ \hline
\end{tabular}
\end{center}
\end{table}

A major benefit of including degenerate triangulations is the reduction of
geometric finite-size effects. An example of this is the volume
dependence of the pseudo-critical cosmological constant $\mu_c(N_4)$
shown in Figure~1. For comparison we show the corresponding
values for combinatorial triangulations. For degenerate triangulations we
observe a rapid convergence to an infinite volume value $\bar{\mu}$;
this is quantified by fits to two different functional forms:
a power-law convergence, 
\vspace{-18pt}
\begin{equation}
\mu_c(N_4) \;=\; \bar{\mu} + \frac{b}{N_4^{\gamma}},  \label{exp}
\end{equation}
\vspace{-3pt}
and a logarithmic divergence, 
\vspace{-3pt}
\begin{equation}
\mu_c(N_4) \;=\; \bar{\mu} + b \log N_4.  \label{super}
\end{equation}
\vspace{-3pt}
The fit parameters and the quality of the fits are shown in Table 1.
Contrary to combinatorial triangulations, for degenerate
triangulations there is no comparison in the quality of the fits;
the latter, corresponding to
a super-exponential growth of the canonical partition function,
is ruled out by a $\chi^2/(\mathrm{d.o.f.}) \approx 117$ for a 
fit to Eq.~(\ref{super}). In
contrast, for combinatorial triangulations it is difficult to 
distinguish between the two scenarios based on the quality
of the fits \cite{bound4d}.
The importance of this
result is that it provides strong numerical
evidence for an exponential bound on the ensemble of degenerate
triangulations as a function of volume. And, 
as $\mathcal{T}_C \in \mathcal{T}_D$, this implies an exponential bound on the number of combinatorial
triangulations as well.

\begin{figure}[t]
\centerline{\epsfig{file=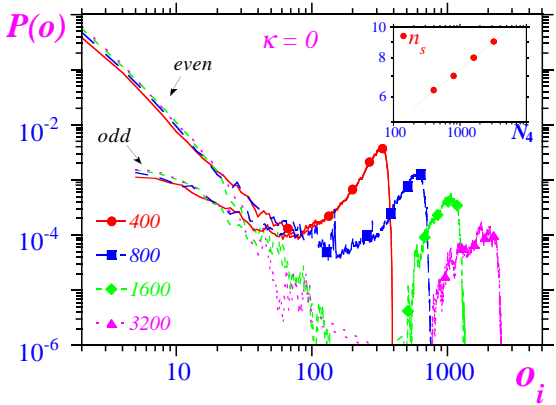,width=3.2in}}
\caption{The (normalized) distributions of vertexe orders for
degenerate triangulations. 
This is for $N_{4}\geq 400$ and $\protect\kappa =0$.
{\sl Insert}: The number of singular vertexes $n_s$.}
\label{fig2}
\end{figure}

As for combinatorial triangulations, in the crumpled phase 
the internal geometry of degenerate triangulations is dominated by a
singular structure.    
The probability distribution of the vertex orders $o_i$
--- the number of 4-simplexes containing the vertex ---
contains an isolated peak in the tail,
indicating singular vertexes (Figure 2).  
However, the distributions $P(o)$ differs in two respects 
from the corresponding ones measured on combinatorial triangulations: 

\begin{mydescription}

\vspace{-6pt}
\item[({\tt i})] 
 The number of singular vertexes
 is larger than two and increases (logarithmically) with the volume,
 i.e.\  a {\it gas} of singular vertexes.

\vspace{-6pt}
\item[({\tt ii})] 
 On each volume the distribution effectively
 separates in two, depending on whether the vertex orders
 are odd or even.

\end{mydescription}
\vspace{-6pt}
It is not clear though, how much significance should be attached to
this observed difference.  Due to the collapsed
nature of the internal geometry it is unlikely that any
sensible continuum limit exist in the crumpled
phase, hence there is no reason to expect identical 
scaling behavior for the two different ensembles.

Additional evidence of a collapsed intrinsic geometry
comes from the (absence of) volume scaling of the simplex-simplex 
correlation function, from which 
we conclude that $d_H = \infty$.

We have also investigated the phase structure of the model for non-zero
values of the inverse Newtons's constant $\kappa$. As for combinatorial
triangulations we observe a phase transition at a value $\kappa_c \approx 1.5
$. For $\kappa > \kappa_c$ the model is in a branched
polymer phase; this we establish by measuring the fractal dimensions $%
d_H$ and the spectral dimension $d_s$ for $\kappa = 2.0$. 
Including measurements on volumes,  
$N_4 = 400$ to 1600 we get $d_H = 1.9(2)$ and $d_s = 1.32(5) $,
in excellent agreement with $d_H = 2$ and $d_s = 4/3$ expected 
for branched polymers. 

\vspace{3pt}
Our results demonstrate that including degenerate triangulations in
simulations of four-dimensional simplicial gravity has many potential
advantages over a model restricted to combinatorial triangulations. 
This agrees with
the same observation previously made in both two and three dimensions. The
chief benefit is the reduction in geometric finite-size effects, mainly due
to an enlarged ensemble --- with a larger triangulation-space
the infinite-volume fractal structure is more easily approximated on
small volumes.  Further work is needed to fully explore the nature
of this ensemble, for now
the most important result we want to emphasize is the
strong numerical evidence for an exponential bound on the
canonical ensemble $\Omega _{N_{4}}$ of degenerate
triangulations.


\end{document}